\documentclass[prd,onecolumn]{revtex4}
\usepackage{dcolumn}
\usepackage{multirow}
\usepackage{graphicx}
\usepackage{amssymb}
\usepackage{bm}
\usepackage{hyperref}
\usepackage{epstopdf}
\usepackage{color}
\usepackage{mathrsfs}
\usepackage{amsmath,amssymb,amsthm}
\usepackage{rotating}
\usepackage{sverb, longtable}
\usepackage{subfigure}
\begin{document}
\title{Exploring New Physics Beyond the Standard Cosmology with Dark Energy Survey Year 1 Data}
\author{Deng Wang}
\email{cstar@sjtu.edu.cn}
\affiliation{National Astronomical Observatories, Chinese Academy of Sciences, Beijing, 100012, China}

\begin{abstract}
With the recent data from the first year of the Dark Energy Survey (DES Y1), we attempt to probe whether there is new physics beyond the standard cosmology and to reconcile the $\sigma_8$ tension between the DES Y1 and Planck datasets in three alternative cosmological models. Combining the galaxy clustering and weak gravitational lensing data from the DES Y1 with other cosmological observations including cosmic microwave background radiation, baryon acoustic oscillations, Type Ia supernovae (SNe Ia) and Planck lensing, we find that: (i) For the interacting dark energy model, we have an updated constraint on the modified matter expansion rate $\epsilon=-0.00004\pm 0.00031$, which indicates that there is no hint of interaction between dark matter and dark energy in the dark sector of the universe; (ii) For the viscous dark energy model, up to now, the strictest constraint on the bulk viscosity coefficient $\eta<0.00152$ at the $2\sigma$ confidence level is obtained; (iii) For the dynamical dark energy model, our constraint on the typical parameter $\beta=-0.04\pm0.12$ implies that there is still no evidence of dynamical dark energy. Meanwhile, we find the $\sigma_8$ tension can be effectively alleviated in the above three alternatives by constraining them with the DES Y1 and Planck datasets separately.

\end{abstract}
\maketitle

\section{Introduction}
During the past two decades, the late-time accelerated expansion of the universe has been confirmed independently by a large number of cosmological probes such as Type Ia supernovae (SNe Ia) \cite{1,2}, baryon acoustic oscillations (BAO) \cite{3} and cosmic microwave background (CMB) radiation \cite{4,5}. To explain this mysterious phenomenon, so far, cosmologists have proposed two main approaches, i.e., modified gravities (MG) and dark energy (DE). The former modifies the standard lagrangian of general relativity (GR) based on some physical-driven mechanism, while the latter introduces an exotic matter source violating the strong energy condition in the Einstein equation in the framework of GR. We would focus on DE issues in this work.

Up to date, making full use of available data, the nature of DE is still unclear and we just know several basic properties of DE: (i) DE is homogeneously
permeated in the universe at cosmological scale and has no the property of clustering unlike dark matter (DM); (ii) a phenomenologically cosmic fluid with an equation of state (EoS) $\omega\approx-1$. To understand the DE phenomenon, the simplest candidate is the so-called $\Lambda$CDM cosmology consisting of cosmological constant and cold DM. This model can successfully explain a wide variety of phenomena, from the origin and evolution of large scale structure to the late-time accelerated universe. Meanwhile, it enjoys a solid motivation that it can be directly interpreted as the vacuum energy density. Nonetheless, this model is not completely perfect and it faces at at least two intractable problems, i.e., coincidence and cosmological constant problems \cite{6}: The former is why energy densities of DM and DE are of the same order of magnitude at present, since their energy densities are so different from each other during the evolutional process of the universe; while the latter suggests the theoretical value for vacuum energy density is far larger than its observed one ($\rho^{th}_{vac}\gg\rho^{obs}_{vac}$), namely the so-called 120-orders-of-magnitude discrepancy that makes the vacuum explanation very confusing. Apart from these problems, in recent several years, two main tensions emerge between high and low redshift probes. On the one hand, the Hubble constant $H_0$ derived indirectly from the Planck datasets \cite{7} under the assumption of $\Lambda$CDM is lower than the direct local measurement from Riess {\it et al.} over the $3\sigma$ confidence level (CL) by using improved SNe Ia calibration techniques \cite{8}. On the other hand, the derived amplitude of the rms density fluctuations $\sigma_8$ from the Planck datasets in the linear regime is higher than the same quantity measured by several low-redshift large scale structure probes including cluster counts, weak lensing and redshift space distortion \cite{9,10}.

Ever since the large scale structure surveys CFHTLenS \cite{11} and KiDS-450 \cite{12}, the $\sigma_8$ tension is recently confirmed, once again, by the first year data release of the DES \cite{13}, which will map 300 million galaxies and tens of thousands of galaxy clusters in five filters ({\it grizY}) over 5000 deg$^2$. The DES collaboration presents cosmological results from a joint analysis of galaxy clustering and weak gravitational lensing by utilizing 1321 deg$^2$ of {\it griz} imaging data from the DES Y1. They find that the DES Y1 best-fitting values for present-day matter density $\Omega_m$ and compound amplitude parameter $S_8$ ($S_8\equiv\sigma_8(\Omega_m/0.3)^{0.5}$) are lower than those from the Planck CMB datasets for $\Lambda$CDM and $\omega$CDM ($\omega$ is EoS of DE), and that the Bayesian evidence indicates that there is no preference between the DES Y1 and Planck datasets under the assumption of $\Lambda$CDM \cite{13}. In light of this $\sigma_8$ tension, the DES collaboration has used several extended models to reconcile it, including non-zero curvature, massless neutrinos, time-varying EoS of DE and MG scenarios \cite{14}. It is interesting that the $\sigma_8$ tension is still not alleviated in the model-independent parameterized MG model. Our motivations are to resolve this $\sigma_8$ tension between DES Y1 and Planck CMB datasets by using three popular physical-driven one-parameter extensions to $\Lambda$CDM, including interacting DE, bulk viscosity DE and dynamical DE, and to explore whether there exists new physics beyond the standard cosmology by combining DES Y1 with other datasets in our numerical analysis.

This work is organized in the following manner. In the next section, we introduce the necessarily analytic formula of three cosmological models. In Section III, we describe four kinds of datasets and statistical techniques used in this work. In Section IV, we present our results from data analysis. The discussions and conclusions are shown in the final section.   

\section{Models}
In this section, we shall describe briefly three alternative models to be constrained by data.
We start with a homogeneous and isotropic universe described by the Friedmann-Robertson-Walker (FRW) metric
\begin{equation}
ds^2=-dt^2+a^2(t)\left[\frac{dr^2}{1-Kr^2}+r^2d\theta^2+r^2sin^2\theta d\phi^2\right],      \label{1}
\end{equation}   
where $a(t)$ and $K$ denotes the scale factor at cosmic time $t$ and the Gaussian curvature of spacetime, respectively. In the context of GR, inserting Eq. \ref{1} into the Einstein equation, one can obtain the so-called Friedmann equations characterizing the dynamics of matter components of the universe as follows
\begin{equation}
H^2=\frac{8\pi G}{3}\Sigma\rho_i,     \label{2}
\end{equation}   
\begin{equation}
\frac{\ddot{a}}{a}=-\frac{4\pi G}{3}\Sigma(\rho_i+3p_i),     \label{3}
\end{equation}   
where $H$ is the Hubble parameter and $\rho_i$ and $p_i$ denote mean energy density and mean pressure of different components in the cosmic pie. Since focusing on the late-time universe, we do not take the contribution from radiation into consideration. Notice that we use the units $8\pi G=c=\hbar=1$ throughout this study. In a flat FRW universe, combining Eqs. \ref{2}-\ref{3}, one can have the dimensionless Hubble parameter (DHP) for the $\Lambda$CDM model
\begin{equation}
E_{\mathrm{\Lambda CDM}}(z)=\left[\Omega_{m}(1+z)^3+1-\Omega_{m}\right]^{\frac{1}{2}}, \label{4}
\end{equation}
where $z$ is the redshift and the EoS of DE has been fixed to be -1 \cite{15}. 

Like DE, the nature of DM is also unclear, therefore, it is reasonable to resolve the coincidence problem by introducing the interaction between DM and DE in the dark sector of the universe. It is rather natural to introduce the interaction in the dark sector by considering a modified matter expansion rate $\epsilon$ in the standard formalism, $\rho_m=\rho_{m0}a^{-3+\epsilon}$, where the free parameter $\epsilon>0$ means that the momentum transfers from DE to DM and vice versa. Subsequently, inserting this basic assumption into Eqs.(2) and (3), the DHP of interacting DE (IDE) model \cite{16} we will analyze is written as
\begin{equation}
E_{\mathrm{IDE}}(z)=\left[\frac{3\Omega_{m}}{3-\epsilon}(1+z)^{3-\epsilon}+1-\frac{3\Omega_{m}}{3-\epsilon}\right]^{\frac{1}{2}}.   \label{5}
\end{equation}
It is worth noting that there are a large number of researches about confronting various IDE models with cosmological observations \cite{a1,a2,a3,a4,a5,a6}. 

Since we do not know whether the nature current DE is particle, it is reasonable to assume that DE is a diluted cosmic fluid permeated in the universe. Based on the thermodynamic concern, there will exist an extra disperse term in the pressure of DE.    
In order to alleviate the $H_0$ and $\sigma_8$ tensions in our previous work, we propose an interesting viscous DE (VDE) model \cite{17}, where the effective pressure of DE is shown as $p_{de}=-\rho_{de}-3\eta H^2$ by adding a bulk viscosity term $-3\eta H^2$ into the standard case. $\rho_{de}$ and $\eta$ denote the energy density of DE and bulk viscosity coefficient, respectively. 
The corresponding DHP of this VDE model is expressed as
\begin{equation}
E_{\mathrm{VDE}}(z)=\left[\frac{\Omega_{m}}{1+\eta}(1+z)^3+(1-\frac{\Omega_{m}}{1+\eta})(1+z)^{-3\eta}\right]^{\frac{1}{2}}.   \label{6}
\end{equation}

Note that the viscosity term $-3\eta H^2$ is added by hand here.

An important and unsettled problem in modern cosmology is whether the DE is a dynamical matter component. In general, one study this problem by modeling the EoS of DE $\omega=\omega(z)$ with various expressions. However, one can also parameterize the DE density to explore this problem. To deal with this problem, we utilize a DE density-parametrization scenario (hereafter we call this model as DDE) \cite{18} $\rho_{de}=\rho_{de0}\left[1+\beta(1-a)\right]$. Substituting this assumption into the Friedmann Eqs.(2) and (3),
the DHP of which can be shown as 
\begin{equation}
E_{\mathrm{DDE}}(z)=\left[\Omega_{m}(1+z)^{3}+(1-\Omega_{m})(1+\beta-\frac{\beta}{1+z})\right]^{\frac{1}{2}},   \label{7}
\end{equation}
where $\beta$ is a typical parameter characterizing this model. One can easily find that when $\epsilon=\eta=\beta=0$, the above three alternatives will reduce to $\Lambda$CDM. These three one-parameter extensions to $\Lambda$CDM have simplicity and elegance to be tested with observations and can easily tell us that whether there is an interaction in the dark sector, bulk viscosity of DE or dynamical hints of DE.

Besides the background evolution of the above three cosmological models, we also consider modifications to linear perturbations of FRW metric. In general, the scalar perturbation of FRW spacetime \cite{19} can be written as
\begin{equation}
ds^2=-(1+2\Phi)dt^2+2a\partial_iBdtdx+a^2\left[(1-2\Psi)\delta_{ij}+2\partial_i\partial_jE\right]dx^idx^j,  \label{8}
\end{equation}
where $\Phi$ and $\Psi$ are linear gravitational potentials. Under the synchronous gauge setting $\Psi=\eta$, $\Phi=B=0$ and $E=-(h+6\eta)/2k^2$, then in $k$-space, the perturbed part of energy-momentum conservation equations will leads to
\begin{equation}
\dot{\delta}=-(1+\omega)(\theta+\frac{\dot{h}}{2})+3(\omega-\frac{\delta p}{\delta\rho})\frac{\dot{a}}{a}\delta, \label{9}
\end{equation}
\begin{equation}
\dot{\theta}=(3\omega-1)\frac{\dot{a}}{a}\theta-\frac{\dot{\omega}}{1+\dot{\omega}}\theta+\frac{k^2\delta}{1+\omega}\frac{\delta p}{\delta\rho}-k^2\sigma, \label{10}
\end{equation}
where $\omega$, $\sigma$, $\delta$ and $\theta$ denote, respectively, the EoSs of cosmic fluids, shear stress, density perturbation and velocity perturbation, and the dot represents the derivative with respect to the conformal time. Here the shear stress $\sigma$ is also called anisotropy stress perturbation \cite{19}. Since the shear stress term $k^2\sigma$ is far smaller than other terms, we neglect it at the 1-order perturbation level. If $k^2\sigma=0$, the stresses of different components are always isotropic, and Eq.(10) will be simplified. These equations are valid for a single uncoupled fluid, or for the net (mass-averaged) $\delta$ and $\theta$ for all fluids. They need to be modified for individual components if the components interact with each other. Furthermore, we modify the perturbations of DE as follows
\begin{equation}
\dot{\delta}_{de}=-(1+\omega_{de})(\theta_{de}+\frac{\dot{h}}{2})-3\dot{\omega}_{de}\frac{\dot{a}}{a}\frac{\theta_{de}}{k^2}+3\frac{\dot{a}}{a}(\omega_{de}-c_s^2)[\delta_{de}+3\frac{\dot{a}}{a}(1+\omega_{de})\frac{\theta_{de}}{k^2}],   \label{11}
\end{equation}
\begin{equation}
\dot{\theta}_{de}=(3c_s^2-1)\frac{\dot{a}}{a}\theta_{de}+\frac{c_s^2}{1+\omega_{de}}k^2\delta_{de}, \label{12}
\end{equation}
where $c_s^2$ and $\omega_{de}$ are the sound speed (SS) in the rest frame and the effective EoS of DE, respectively. The definition of sound speed is $c_s^2=\delta P/\delta\rho=\omega$, where $P$ and $\omega$ denote the pressure and EoS of each individual component, respectively. Note that $c_s^2<0$ is unphysical. In principle, $c_s^2$ should be greater than or equal 0.  For photons $\omega=1/3$ and for baryons $\omega\approx0$, therefore, $c_s^2=1/3$ for photons and $c_s^2\approx0$ for baryons. For the DE component, generally, for simplicity, we choose $c_s^2=1$.  Meanwhile, we adopt $\sigma=0$. It is noteworthy that the effective EoSs of DE of IDE, VDE and DDE models can be shown as, respectively, $-1+\frac{(1+z)^{3-\epsilon}-(1+z)^{3}}{\frac{3}{3-\epsilon}(1+z)^{3-\epsilon}-(1+z)^{3}+\frac{\tilde{\Omega_{\Lambda}}}{\Omega_{m}}}$ \cite{16}, $-1-\eta-\frac{\eta\Omega_{m}(1+z)^3}{(1-\frac{1-\eta}{1+\eta}\Omega_{m})(1+z)^{-3\eta}-\frac{\eta}{1+\eta}\Omega_{m}(1+z)^3}$ \cite{17} and $-1+\frac{\beta}{3[(1+\beta)(1+z)-\beta]}$ \cite{18}, where $\tilde{\Omega_{\Lambda}}$ the dimensionless ground state value of vacuum.

\begin{table}[h!]
\renewcommand\arraystretch{1.3}

\caption{The $1\sigma$ confidence intervals from marginalized constraints on model parameters of IDE, VDE and DDE are presented by using the datasets D, C, CBSL and DCBSL, respectively. Here we also quote the $2\sigma$ limits of bulk viscosity coefficient $\eta$ and dynamic parameter $\beta$. }
\label{t1}
\begin{tabular} { l l c c c c c c c }

\hline
\hline
Models & Data & $\Omega_m$ & $\sigma_8$ & $n_s$  & $\epsilon$ & $\eta$ & $\beta$ \\
\hline

IDE  & D  & $0.273^{+0.040}_{-0.035} $   & $0.837\pm0.081$ & $\star$  & $0.029\pm 0.048$  &  $\star$ &   $\star$                                        \\
IDE  & C  & $0.302\pm0.015 $   & $0.805\pm0.018$ & $0.9697\pm0.0068$  & $-0.0013\pm 0.0038      $  & $\star$  &   $\star$                                        \\
IDE  & CBSL  & $0.3216\pm 0.0072$   & $0.810\pm0.016$ & $0.9569\pm0.0040 $  & $-0.00006\pm 0.00031      $  & $\star$  & $\star$                                          \\
IDE  & DCBSL  & $0.309\pm0.006$   & $0.792^{+0.015}_{-0.013}$ & $0.9608\pm0.0040 $  & $-0.00004\pm 0.00031       $  & $\star$  &  $\star$                                         \\
\hline
VDE  & D  & $0.283\pm0.029$   & $0.798\pm 0.052$ & $\star$  & $\star$ & $<0.879 \, (2\sigma)$  &  $\star$                                         \\
VDE  & C  & $0.296^{+0.014}_{-0.016}$  & $0.804\pm 0.018$ & $0.9738\pm0.0072 $  & $\star$  & $<0.185 \, (2\sigma)$  &    $\star$                                       \\
VDE  & CBSL  & $0.307^{+0.013}_{-0.012}$   & $0.816\pm 0.011           $ & $0.9671\pm0.0077 $  & $\star$  & $<0.00348 \, (2\sigma)$  &   $\star$                                        \\
VDE  & DCBSL  & $0.301\pm0.010$   & $0.807\pm 0.006           $ & $0.9687\pm0.0079 $  & $\star$ & $<0.00152 \, (2\sigma)$  &   $\star$                                        \\
\hline
DDE  & D  & $0.275^{+0.036}_{-0.040}      $   & $0.853\pm 0.077           $ & $\star$  & $\star$  & $\star$  & $>-0.684 \, (2\sigma)$                                          \\
DDE  & C  & $0.320\pm 0.021           $   & $0.801^{+0.041}_{-0.029}          $ & $0.9604\pm0.0041 $  & $\star$  &  $\star$ & $-0.25^{+0.28}_{-0.30}$                                          \\
DDE  & CBSL  & $0.315\pm 0.008           $   & $0.830^{+0.016}_{-0.013}           $ & $0.9534\pm0.0048 $  & $\star$  & $\star$  & $-0.23^{+0.12}_{-0.14}$                                          \\
DDE  & DCBSL  & $0.305^{+0.008}_{-0.007}  $   & $0.809\pm 0.010           $ & $0.9622\pm0.0045$ & $\star$ & $\star$  & $-0.04\pm 0.12$                                          \\

\hline
\hline
\end{tabular}
\end{table}

\begin{figure}[htbp]
\centering

\subfigure[\quad IDE]{
\begin{minipage}[htbp]{0.5\linewidth}
\centering
\includegraphics[width=10cm, height=10cm]{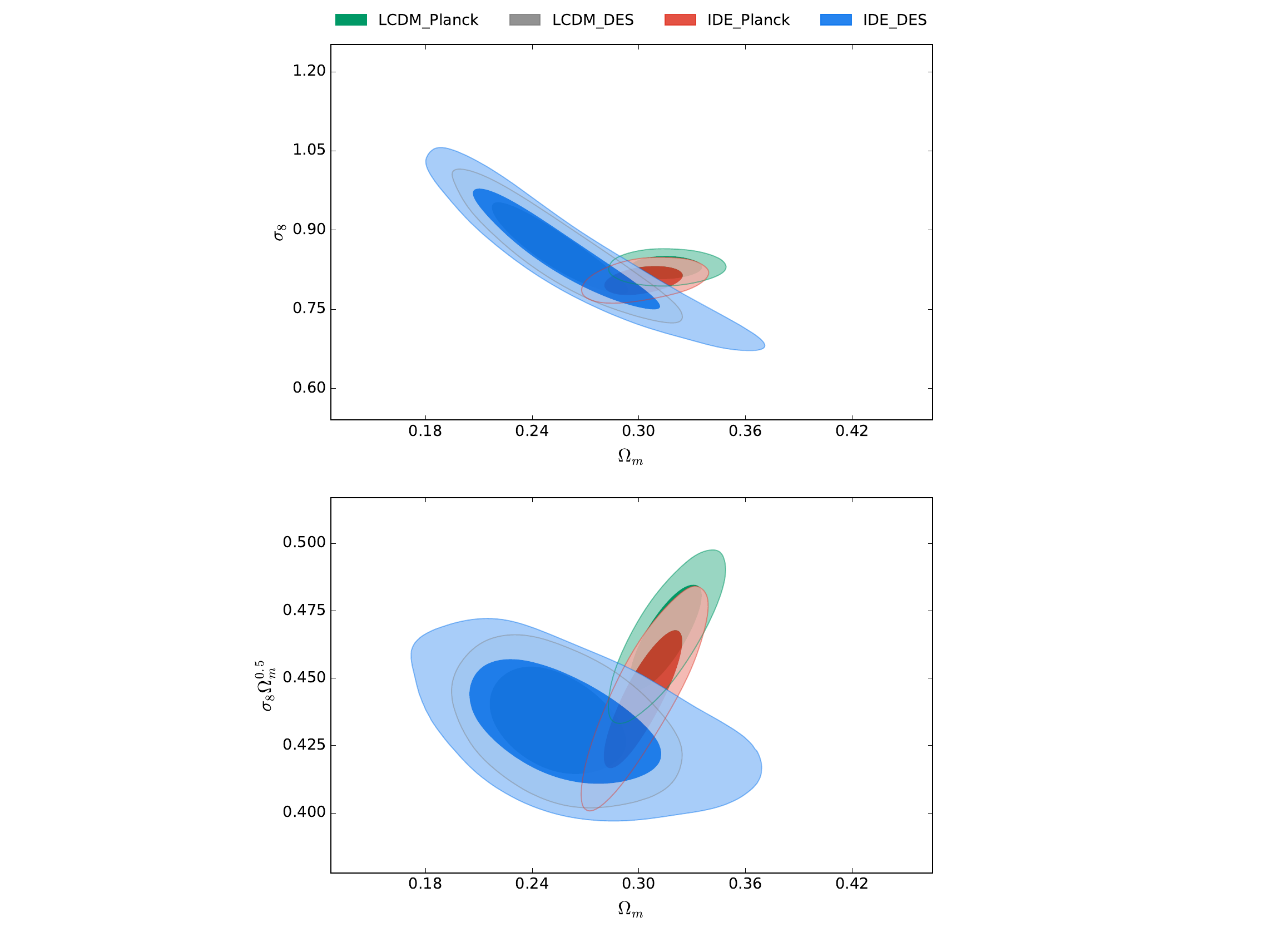}
\end{minipage}%
}%
\\                 
\subfigure[\quad VDE]{
\begin{minipage}[htbp]{0.5\linewidth}
\centering
\includegraphics[width=10cm, height=10cm]{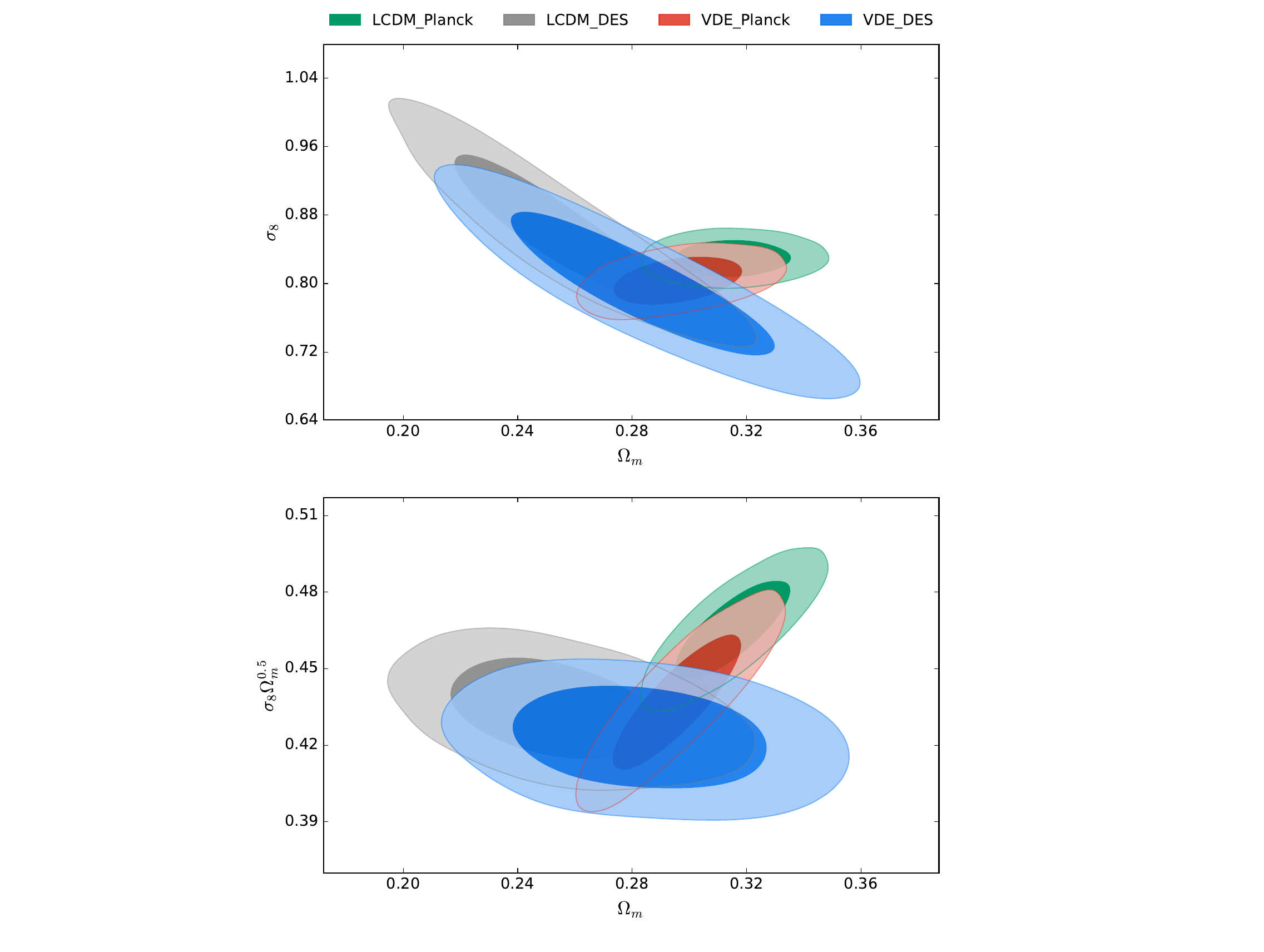}
\end{minipage}
}%
\subfigure[\quad DDE]{
\begin{minipage}[htbp]{0.5\linewidth}
\centering
\includegraphics[width=10cm, height=10cm]{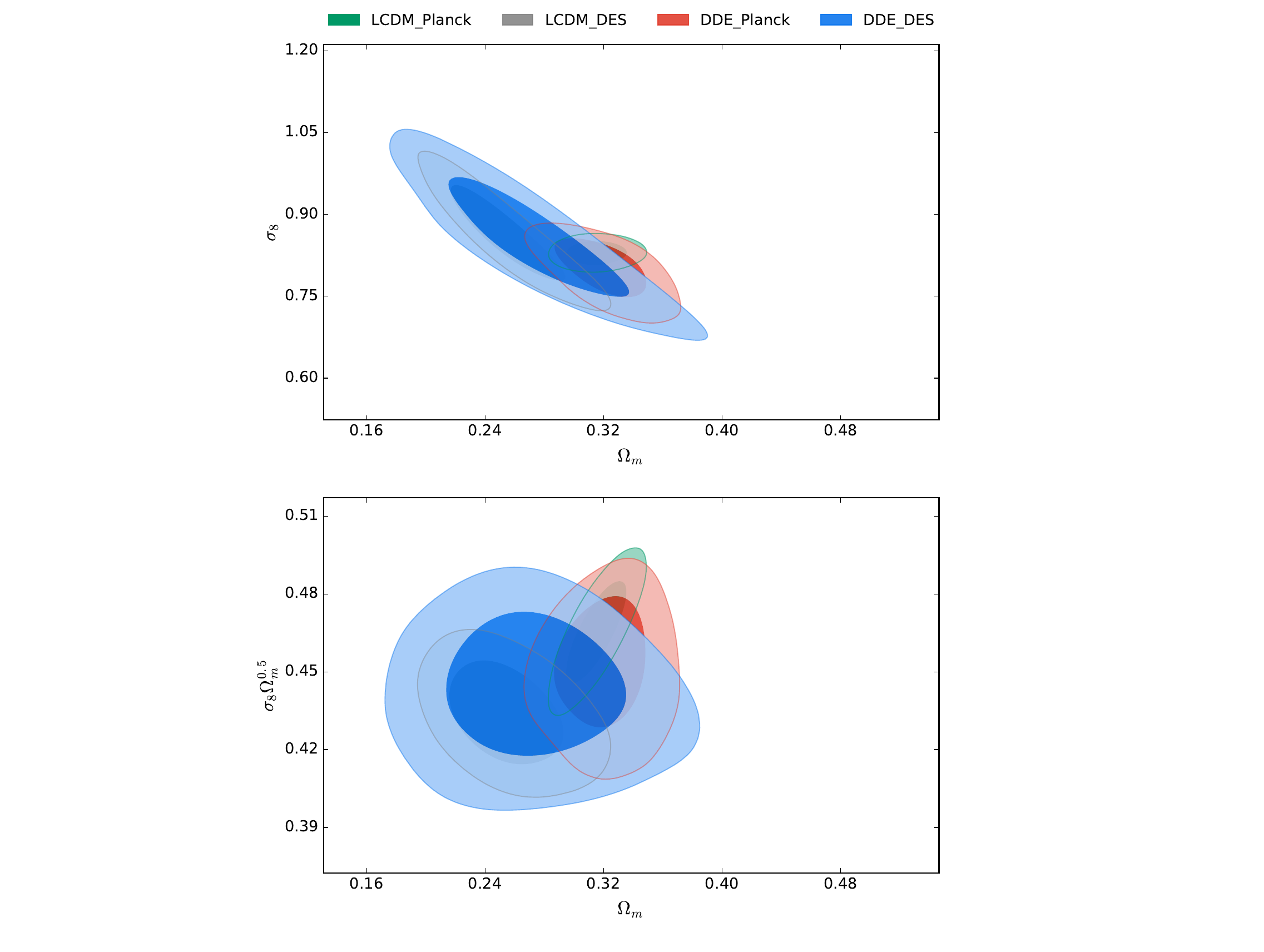}
\end{minipage}
}%

\centering
\caption{The $1\sigma$ and $2\sigma$ marginalized contours of IDE, VDE and DDE models are presented by using the datasets D (blue) and C (red) in the planes of $\Omega_m-\sigma_8$ and $\Omega_m-\sigma_8\Omega_m^{0.5}$, respectively. As a comparison, the $1\sigma$ and $2\sigma$ marginalized contours of $\Lambda$CDM model using the datasets C (green) and D (grey) are also shown, respectively. }\label{f1}
\end{figure}


\begin{figure}
\centering
\includegraphics[scale=0.5]{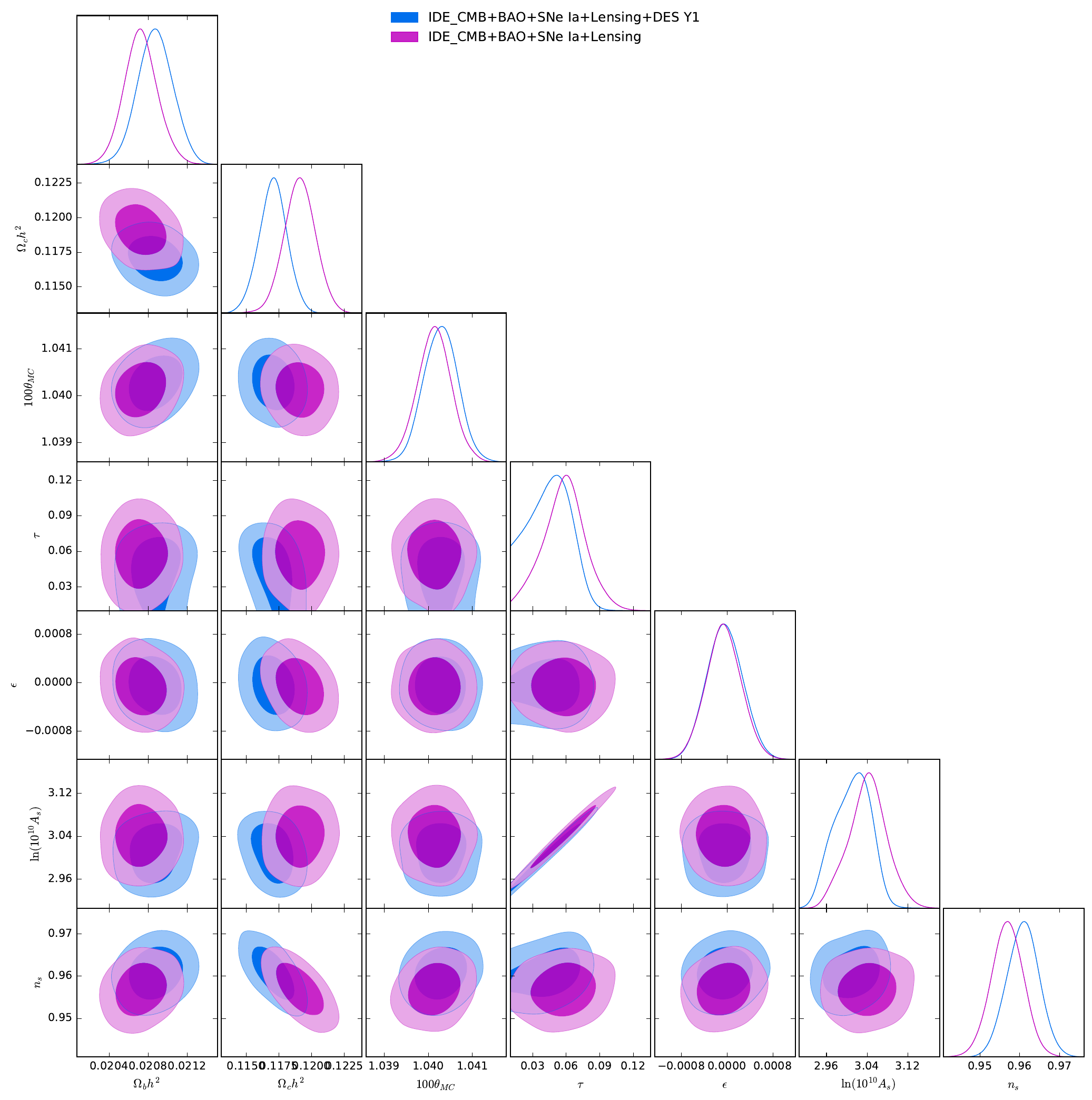}
\caption{The $1\sigma$ and $2\sigma$ marginalized constraints on the cosmological parameters of IDE model are presented by using the combined datasets CBSL (magenta) and DCBSL (blue), respectively. }\label{f3}
\end{figure}

\begin{figure}
\centering
\includegraphics[scale=0.5]{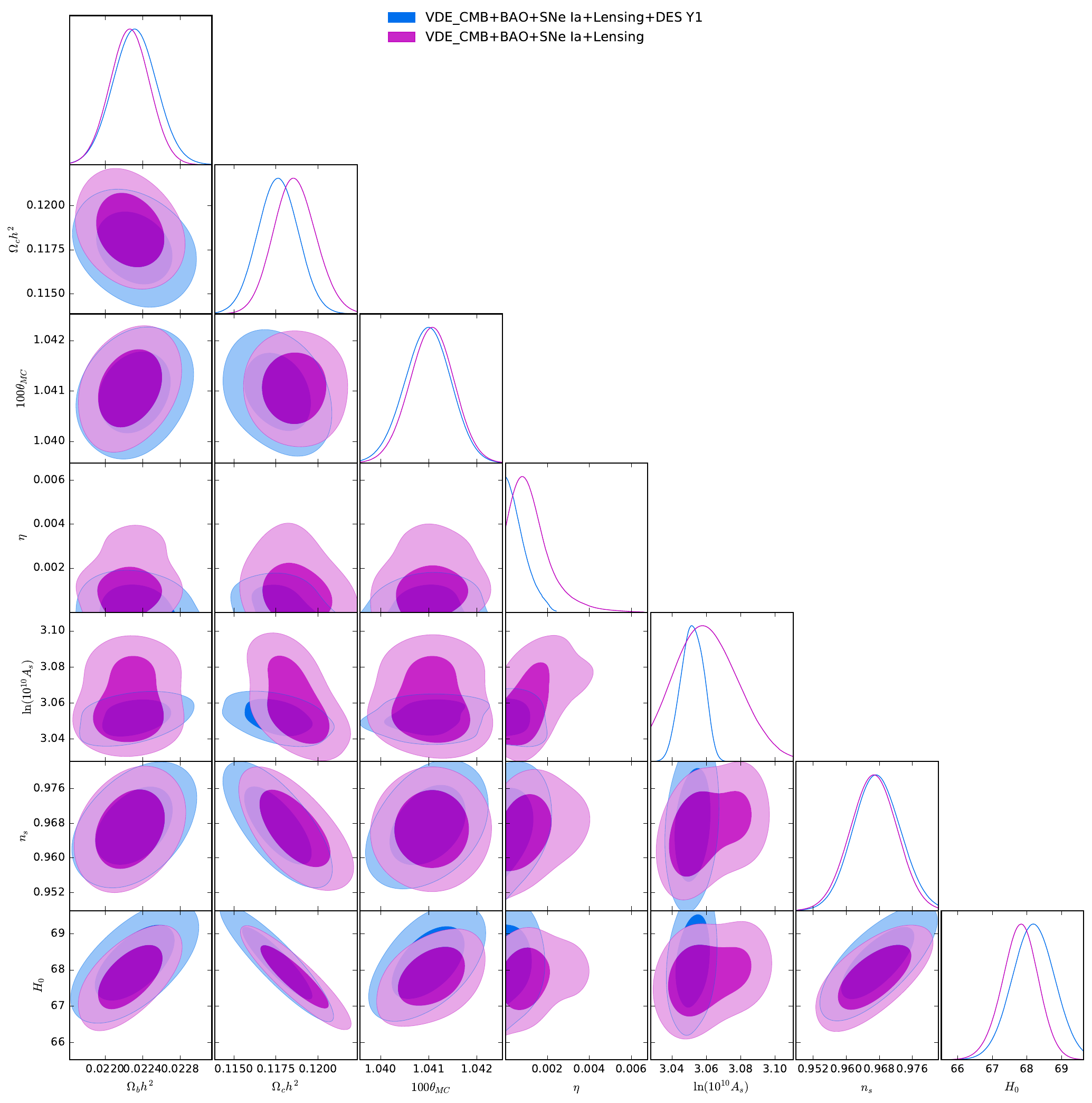}
\caption{The $1\sigma$ and $2\sigma$ marginalized constraints on the cosmological parameters of VDE model are presented by using the combined datasets CBSL (magenta) and DCBSL (blue), respectively.  }\label{f4}
\end{figure}
\begin{figure}
\centering
\includegraphics[scale=0.5]{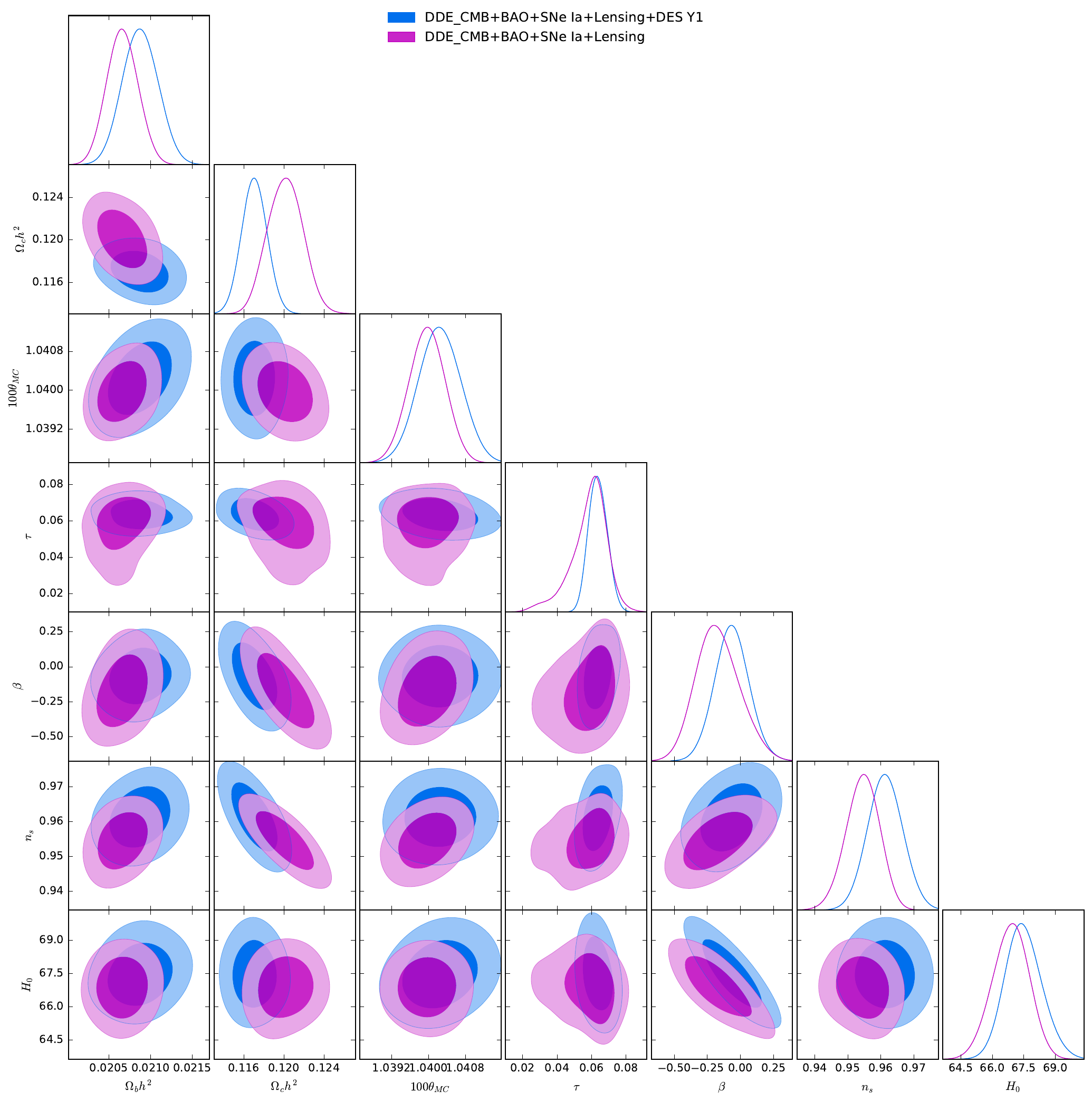}
\caption{The $1\sigma$ and $2\sigma$ marginalized constraints on the cosmological parameters of DDE model are presented by using the combined datasets CBSL (magenta) and DCBSL (blue), respectively.  }\label{f5}
\end{figure}

\section{Data and Statistics}
To perform the observational constraints on the above three alternatives better by using the latest cosmological observations, we shall write their parameter spaces as follows
\begin{equation}
\mathbf{P_{\mathrm{\Lambda CDM}}}=\{\Omega_bh^2, \quad \Omega_ch^2, \quad 100\theta_{MC}, \quad \tau, \quad  \mathrm{ln}(10^{10}A_s), \quad  n_s \},   \label{13}
\end{equation}

\begin{equation}
\mathbf{P_{\mathrm{IDE}}}=\{\Omega_bh^2, \quad \Omega_ch^2, \quad 100\theta_{MC}, \quad \tau, \quad  \mathrm{ln}(10^{10}A_s), \quad  n_s, \quad \epsilon \},   \label{15}
\end{equation}
\begin{equation}
\mathbf{P_{\mathrm{VDE}}}=\{\Omega_bh^2, \quad \Omega_ch^2, \quad 100\theta_{MC}, \quad \tau, \quad  \mathrm{ln}(10^{10}A_s), \quad  n_s, \quad \eta \},   \label{16}
\end{equation}
\begin{equation}
\mathbf{P_{\mathrm{DDE}}}=\{\Omega_bh^2, \quad \Omega_ch^2, \quad 100\theta_{MC}, \quad \tau, \quad  \mathrm{ln}(10^{10}A_s), \quad  n_s, \quad \beta \},   \label{17}
\end{equation}
where $\Omega_bh^2$ and $\Omega_ch^2$ denote the present baryon and CDM densities, $\theta_{MC}$ is the ratio between the angular diameter distance and sound horizon at the redshift of last scattering $z_\star$, $\tau$ is the optical depth due to reionization, $\mathrm{ln}(10^{10}A_s)$ the amplitude of primordial power spectrum at the pivot scale $K_0=0.05$ Mpc$^{-1}$, $n_s$ the scalar spectral index, $\epsilon$ the modified matter expansion rate, $\eta$ the bulk viscosity coefficient and $\beta$ the typical parameter of our DDE model. Here $h\equiv H_0/(100$ km s$^{-1}$ Mpc$^{-1})$.

In this work, we shall use several standard cosmic probes to explore whether there exists new physics beyond the standard cosmological model, and try to check that whether the $\sigma_8$ tension between Planck CMB and DES Y1 datasets can be appropriately reconciled in the above alternatives. 

The specific probes from DES Y1 \cite{13} used in this work include the following three 2-point correlation functions:

{\it Galaxy clustering}: Generally, the homogeneity of matter distribution in the universe can be traced via galaxies distribution. The overabundance of pairs at angular separation $\theta$ in a random distribution, $\omega(\theta)$, is one of the most convenient ways to measure galaxy clustering. It quantifies the scale dependence and strength of galaxy clustering, and consequently affects the matter clustering \cite{20}.

{\it Cosmic shear}: The 2-point statistics characterize shapes of galaxies are very complex, since they are products of components of a spin-2 tensor. Hence, it is convenient to extract information from a galaxy survey by using a pair of 2-point correlation functions $\xi_+(\theta)$ and $\xi_-(\theta)$, which denote the sum and difference of products of tangential and cross components of the shear, measured with respect to the line connecting each pair of galaxies \cite{21}. 

{\it Galaxy-galaxy lensing}: The characteristic distortion of source galaxy shapes is from mass associated with foreground lenses. This distortion is the mean tangential ellipticity of source galaxy shapes around lens galaxy positions for each pair of redshift bins and also named as the tangential shear, $\gamma_t(\theta)$ \cite{22}.

More detailed information and possible systematic effects of the DES Y1 data can be found in \cite{20,21,22}. We will refer to this dataset as `` D '' later.  

The external cosmic probes included in this work shall be shown as follows:

{\it CMB}: The Planck satellite has measured many aspects of formation and evolution of the universe such as matter components, topology and large scale structure effects. We use Planck-2015 CMB temperature and polarization data including likelihoods of temperature at $30\leqslant \ell\leqslant 2500$ and the low-$\ell$ temperature and polarization likelihoods at $2\leqslant \ell\leqslant 29$, i.e., TT$+$lowP \cite{15}. This dataset is denoted as `` C ''.

{\it BAO}: The BAO is a very clean observation to probe the evolution of the universe, which are unaffected by errors in the nonlinear evolution of the matter density field and other systematic errors which may affect other observations. Measuring the position of these oscillations in the matter power spectrum at different redshifts can place constraints on the expansion history of the universe after decoupling and break degeneracies of parameters better. Here we employ the BOSS DR12 dataset at three effective redshifts $z_{eff}=$ 0.38, 0.51 and 0.61 \cite{23}, the 6dFGS sample at $z_{eff}=$ 0.106 \cite{24} and the SDSS-MGS one at $z_{eff}=$ 0.15 \cite{25}. We refer to this dataset as `` B ''.

{\it SNe Ia}: SNe Ia is a powerful distance indicator to probe the expansion history of the universe, especially, the EoS of DE. To implement numerical analysis, we utilize the largest SNe Ia `` Pantheon '' sample to date, which integrates the SNe Ia data from the Pan-STARRS1, SNLS, SDSS, low-z and HST and consists of 1049 spectroscopically confirmed SNe Ia in the redshift range $z \in [0.01, 2.3]$ (see \cite{26} for more details). Hereafter we denote this dataset as `` S ''.

{\it Lensing}: As a complementary probe, we also make use of Planck-2015 CMB lensing measurements from temperature only \cite{27}, which act like an extra narrow and very high redshift source sample are measured from, are higher-order correlations in the temperature field, and have given high-quality measurement with a 2.5$\%$ constraint on the amplitude of lensing potential power spectrum. We shall refer to this dataset as `` L ''.

To acquire the posterior probability density distributions of model parameters, we modify equations governing the background evolution and perturbations of the universe in the public package CosmoMC and CAMB \cite{28,29}. To implement the standard Bayesian analysis, we choose the following prior ranges for different model parameters: $\Omega_bh^2 \in [0.005, 0.1]$, $\Omega_ch^2 \in [0.001, 0.99]$, $100\theta_{MC} \in [0.5, 10]$, $\tau \in [0.01, 0.8]$, $\mathrm{ln}(10^{10}A_s) \in [2, 4]$, $n_s \in [0.8, 1.2]$, $\epsilon \in [-0.3, 0.3]$, $\eta \in [0, 1]$ and $\beta \in [-3, 3]$. To compare the impacts of different datasets on constraining typical model parameters, we take four kinds of data combinations: (i) C; (ii) D; (iii) CBSL; (iv) DCBSL.

It seems that a negative viscosity does not have a simple physical interpretation. However, it is a formal quantity. In a cosmological model with negative bulk viscosity, this viscosity contributes with an attractive gravity, and thus tends to decrease the cosmic expansion. Even if the cosmic fluid starts within the phantom region, it becomes possible, if the negative viscosity is large enough, to abandon the future singularity. As a consequence, the prior $\eta \in [0, 1]$ used here is relatively restrictive.
The concept of negative viscosity is related to the concept of negative temperature. One can find further explanations of negative temperature in Ref.\cite{a7}. For simplicity, we assume the positive viscosity, which leads to the accelerated expansion of the universe.  

\section{Results}

In light of different kinds of datasets D, C, CBSL and DCBSL, our numerical analysis results are presented in Tab. \ref{t1}, which includes mean values with $1\sigma$ errors and $2\sigma$ limits of different model parameters of the above three alternatives from Markov Chain Monte Carlo analysis. The marginalized 1-dimensional and 2-dimensional posterior distributions of cosmological parameters of these three models are also exhibited in Figs. \ref{f1}-\ref{f5}. In Fig. \ref{f1}, one can easily find that the $\sigma_8$ tension between DES Y1 and Planck datasets are clearly alleviated in IDE, VDE and DDE models. As a comparison, we also show the corresponding $\Omega_m-\sigma_8\Omega_m^{0.5}$ contours and the same conclusion is obtained.

For the IDE model, in the DES Y1-only case, we obtain a relatively loose constraint on the modified matter expansion rate $\epsilon=0.029\pm 0.048$ by comparing with Planck's constraint $\epsilon=-0.0013\pm 0.0038$. By use of the combined dataset CBSL and DCBSL, we have improved constraints on the IDE parameter $\epsilon=-0.00006\pm 0.00031$ and $-0.00004\pm 0.00031$, the cental values of which are one order of magnitude lower than those obtained in our previous work \cite{30,31}. Meanwhile, there is still no hint of interaction in the dark sector of the universe. For the VDE model, the $2\sigma$ upper bound on the bulk viscosity coefficient $\eta<0.879$ in the DES Y1-only case is much larger than that in the Planck-only case ($\eta<0.185$). Utilizing the data combination of CBSL, we obtain a smaller $2\sigma$ bound $\eta<0.00348$ than DES Y1 and Planck data. If combining DES Y1 with CBSL, we can have a tighter $2\sigma$ upper limit $\eta<0.00152$ than $\eta<0.00217$ obtained in our previous work \cite{17}, where we have no use of DES Y1 data. For the DDE model, since the DES Y1 data is insensitive to the background quantity of DDE model, it only gives a $2\sigma$ lower bound on the DDE parameter $\beta>-0.684$. If adding Planck data into BSL, we have a better constraint $\beta=-0.23^{+0.12}_{-0.14}$. However, this indicates a $1.92\sigma$ signal lower than the prediction of standard cosmology. Furthermore, when combing DES Y1 with CBSL, we obtain $\beta=-0.04\pm0.12$ and the above anomaly disappears.
 
Comparing the constraining results from CBSL and those from DCBSL for three alternatives, we find that the addition of DES Y1 data could lead to a larger CDM density ratio $\Omega_{c}h^2$ and a smaller baryon density ratio $\Omega_{b}h^2$, and except for VDE, the values of scalar spectral index $n_s$ shift towards the right in the left two models (see also Figs. \ref{f3}-\ref{f5}). Interestingly, in Fig. \ref{f5}, we also observe that the typical parameter $\beta$ of DDE is anti-correlated with $H_0$ and $\Omega_{c}h^2$. This indicates that increasing $\beta$ would lead to decreasing Hubble constant and CDM density in the late-time universe. Based on current data, two model parameters $\epsilon$ and $\eta$ are still highly degenerated with other cosmological parameters.
We need more high-quality data to break degeneracies between parameters and explore their correlations.

The most direct way to study whether a cosmological model can relieve the $H_0$ tension is constraining it with the Planck CMB data, and then comparing its $H_0$ value with that from the local measurement $H_0=74.03\pm1.42$ km s$^{-1}$ Mpc$^{-1}$  by the HST project \cite{a8}. Using CMB data, we obtain the global $H_0=68.2\pm1.1$, $68.5\pm1.2$ and $68.0^{+1.3}_{-1.4}$ km s$^{-1}$ Mpc$^{-1}$ for IDE, VDE and DDE models, respectively. We find that the $H_0$ tension can be, respectively, alleviated from $4.4\sigma$ to $3.2\sigma$, $2.9\sigma$ and $3.1\sigma$ in these three models.

\section{Discussions and conclusions}
With recent DES Y1 data release, we are motivated by exploring whether there is new physics beyond the $\Lambda$CDM model and alleviating the small $\sigma_8$ tension (see also Fig. 10 in \cite{13}) between the DES Y1 and Planck temperature and polarization datasets by using three alternative cosmological models. 

We find that the $\sigma_8$ tension can be apparently resolved in the IDE, VDE and DDE models.Using the data combination of DCBSL, we have an updated constraint on the IDE parameter $-0.00004\pm 0.00031$, the cental value of which are one order of magnitude lower than that obtained in our previous work \cite{30,31}. This indicates that there is still no evidence of interaction between DM and DE in the dark sector of the universe. For the VDE model, using the combined datasets DCBSL, we obtain a tighter $2\sigma$ upper bound on the bulk viscosity coefficient $\eta<0.00152$ than $\eta<0.00217$ obtained in our previous work \cite{17}.
For the DDE model, adding the DES Y1 data into CBSL, we have the typical parameter of DDE $\beta=-0.04\pm0.12$ at the $1\sigma$ CL, which implies that there does not exist hint of DDE in the universe.

From Tab. \ref{t1}, we can easily see that the scale invariance of primordial power spectrum ($n_s=1$) is still strongly disfavored by current cosmological observations in the above three models. Using the data combinations CBSL and DCBSL, we find the anti-correlations between the DDE parameter $\beta$ and $H_0$ and $\Omega_{c}h^2$, which means that increasing $\beta$ could lead to smaller cosmic expansion rate $H_0$ and CDM density ratio. In light of currently available data, the degeneracies between two model parameters $\epsilon$ and $\eta$ and other ones cannot be well broken.

In light of the constraints presented here, we find that the $H_0$ tension can be partly alleviated in these three models, but can not be solved. We will address carefully this issue in the forthcoming study.

A mature modified gravity theory should have a screening mechanism, which says modified gravity works in underdense regions and reduces to GR in overdense regions. Generally, the local gravity tests require that modified gravity comes back to GR. Therefore, it is hardly possible to detect the fifth force at local scales, e.g., solar systems. These three DE models are based on GR. Since modified gravity and DE are equivalent in the explanation of cosmic acceleration at cosmological scales, they are highly degenerated and we think one may not detect the signal of exotic DE force with local gravity measurements. 

Since the DES collaboration just reports their first year data release which is not good enough to constrain some model parameters with high accuracy, we expect that, by combining their follow-up data with other updated observations, we can explore whether there is any signal beyond the standard cosmological model and go a further step to break the degeneracies between parameters.

\section{Acknowledgements}
We thank Liang Gao, Jie Wang, Qi Guo, Yun Chen, Haonan Zheng, Huijie Hu, Kai Zhu and Hang Yang for useful discussions. This work is supported by the Ministry of Science and Technology of China under Grant No.2017YFB0203300, National Nature Science Foundation of China under Grants No.11988101 and No.11851301.

\end{document}